\documentclass{optica-article}

\journal{opticajournal} 

\articletype{Research Article}

\usepackage{lineno}
\usepackage{subcaption}
\usepackage{amsmath}
\usepackage{array}
\usepackage{booktabs} 
\usepackage[utf8]{inputenc}
\usepackage{cite}

\begin{document}

\title{Advanced pure tilt actuator for testing tilt-to-length coupling in space-based gravitational wave detection}

\author{XIANG LIN, QI XIA, PENG QIU, YU-RONG LIANG and HAO YAN\authormark{*}}

\address{MOE Key Laboratory of Fundamental Physical Quantities Measurement \& Hubei Key Laboratory of Gravitation and Quantum Physics, PGMF and School of Physics, Huazhong University of Science and Technology, Wuhan 430074, China}

\email{\authormark{*}yanhao2022@hust.edu.cn} 


\begin{abstract*} 
Tilt-to-length (TTL) coupling, caused by the jitter of test masses or satellites, is a significant noise source in space-based gravitational wave detection. Calibrating and suppressing TTL coupling noise at the sub-nanometer level is essential. One main challenge in current ground-based TTL coupling testing is the residual translational movement of the tilt actuator. This paper introduces the development of an advanced pure tilt actuator (APTA) specifically designed for testing TTL coupling. The APTA provides precise tilt motion and is monitored by a four-beam interferometer, which measures the displacement of attached array pyramids. We present a detailed theoretical model and experimental setup. Experimental results demonstrate that this optical test bed, equipped with the APTA, can achieve subnanometer-level TTL coupling calibration. In addition, a typical heterodyne interferometer was tested using the APTA test bed. Comparative testing demonstrated that the imaging system is capable of effectively suppressing TTL coupling errors. The TTL coupling coefficients were reduced from over $\pm 30 \, \mu\text{m/rad}$ to within $\pm 5 \, \mu\text{m/rad}$ across a range of $\pm 200 \, \mu\text{rad}$, meeting the preliminary requirements for the TianQin mission. This APTA test platform has the potential to be widely utilized for ground-based TTL coupling inspection.\\
\textcopyright2024 Optica Publishing Group under the terms of the \href{https://www.osapublishing.org/submit/research-articles/authors/open-access-publishing-agreement}{Optica Publishing Group Open Access Publishing Agreement}

\end{abstract*}

\section{Introduction}
In space gravitational wave detection missions such as TianQin\cite{gong2021concepts,luo2016tianqin}, the Laser Interferometer Space Antenna (LISA) \cite{danzmann2003lisa,jennrich2009lisa}, and Taiji \cite{hu2017taiji,luo2020brief}, the demand for displacement detection sensitivity is extremely high. Measurement optical path changes caused by tilt jitter, known as TTL coupling, have become a significant potential noise source, severely affecting the accuracy of displacement measurements along sensitive axes. TTL coupling occurs in both test mass interferometers and long-arm interferometers. The TianQin mission, a space-based gravitational wave detection project in geocentric orbit, constructs a large-scale laser interferometer with an arm length of approximately $\mathrm{1.7 \times 10^{5}\,km}$ . It aims to measure inter-satellite displacement changes with a sensitivity limit of $\mathrm{1\,pm/Hz^{1/2}}$ at 6 mHz [1]. This requirement imposes stringent demands on the suppression of various noise sources, particularly TTL coupling noise. To achieve this goal, the TTL coupling must be minimized to less than $\mathrm{10 \mu m/rad}$ for tilts of $10 \mathrm{nrad/Hz^{1/2}}$ at 6mHz.

Reaching this goal directly is almost impossible, but we can achieve it through several steps. First, we should minimize misalignment error as much as possible during design and installation. However, it is not feasible to perfectly align all components in the system using this method alone, so we must meet the required benchmarks as closely as possible at this phase. Next, we need to calibrate the misalignment. The beam alignment mechanism (BAM), composed of a pair of parallel glass windows, can adjust the transversal position of the beam \cite{brzozowski2022lisa}. The fast steering mirror can correct the beam's angular direction. These two devices are integrated into the system for compensatory adjustments to minimize alignment errors after installation. Finally, misalignments can occur during launch, transfer to orbit, and over time. The remaining TTL noise needs to be subtracted in ground-based post-processing by measuring angular jitter \cite{Paczkowski2022,Armano2023,Hartig2023}, ensuring continuous compliance with gravitational wave detection requirements. By following the three steps, the TTL coupling can be suppressed to the level below our required threshold. However, before achieving this, a clear understanding of the TTL coupling formation mechanism is essential, along with the design and construction of ground-based experimental testing setups.

TTL coupling can arise from various mechanisms, broadly categorized into geometric and non-geometric types \cite{hartig2022geometric,hartig2023non}. In the test mass (TM) interferometer, geometric TTL coupling occurs when the tilt of the TM increases the optical path length from the origin of the measuring light to the detector surface, a phenomenon known as the light lever effect. Misalignment between the TM's center of mass and the reflection point of the measurement laser can also result in a piston effect. Additionally, changes in the angle of the incident measurement light on a plane transmission mirror can alter the optical path along the sensitive axis. Non-geometric TTL coupling involves factors such as the sizes and positions of the beam waists of the two interfering beams, which can affect the readout of the interference signal. Wavefront errors, including higher-order modes and wavefront distortions, modify the wavefront curvature of the interfering beams, potentially impacting the signal readout and inducing TTL coupling errors.  To date, research in this area has been limited. Beyond the interference light itself, the characteristics and geometry of the optoelectronic detector—a critical component in detecting interference signals—directly influence signal readout. Factors such as the detector’s area, slit dimensions, and electrical non-uniformity still need to be accurately characterized.

Ground-based experimental testing on TTL coupling is crucial for the development of space laser interferometers. However, a challenge in experimental testing setups is how to generate pure rotation without introducing parasitic displacement along the sensitive axis, as such displacement would affect the ultimate precise calibration of TTL coupling. Various methods have been proposed in numerous studies, such as using the characteristic of identical optical parameters of two interference beams in a zero-difference interferometer, monitoring and deducting the parasitic displacement of the actuator with a large single element photo diode (SEPD), successfully calibrating the TTL coupling of the device, and verifying that a double lens imaging system can suppress TTL coupling \cite{Sonke2016}. To realistically simulate the LISA optical platform, the LISA team constructed an optical testing platform using a silicate bonding technique, successfully deducing the parasitic displacement of the actuator with technologies such as phase locking between multiple beams \cite{Tröbs2018}. This device can simultaneously simulate TM interferometers and interstellar interferometers \cite{Chwalla2016}. However, this device has a complex optical path structure, and since most optical components are glued on the substrate, it is not easily modifiable or adjustable to test different settings, such as different distances or beam characteristics. To make the TTL coupling calibration device more flexible, the LISA team developed the advanced tilt actuator (ATA), taking advantage of the characteristic that a conical prism is insensitive to changes in the angle of incident light, effectively suppressing the parasitic displacement produced by the actuator \cite{lee2021development}. The ATA was designed and built as an independent experimental platform; hence it can be used not only for a specific TTL coupling experiment but also for other experiments requiring three-degree-of-freedom (DOF) motion of the ATA.

Based on our research on experimental testing setups for TTL coupling, we noticed that the pyramid produces a lateral displacement in the incoming light during rotation, thus affecting the final readout of the interference signal and thus influencing the calibration of the TTL coupling. To address this issue, we designed a self-aligning four-beam interferometer \cite{lin2024compact}. The feature of this device is that the conical prism does not cause lateral displacement of the incoming light when tilted. The device is constructed using a UV adhesive \cite{lin2023construction}, providing compactness and significantly improved displacement measurement sensitivity. This makes it potentially applicable for high-precision TTL coupling experiments in the future.

The paper is organized as follows. Section 2 provides an overview of the mechanisms of TTL coupling formation, which helps us to gain a clearer understanding of TTL coupling. Section 3 details the working principles, optical design, and error analysis of the ATPA, and how this device is applied to ground-based experimental calibration of TTL coupling. Section 4 presents the ground-based experimental testing setup for TTL coupling, demonstrates the calibration of system TTL coupling, and shows the suppression of TTL coupling using a confocal imaging system. Section 5 contains the discussion.

\section{TTL coupling and imaging system}
\subsection{Mechanism of TTL coupling}
TTL coupling in laser interferometers is a complex effect. Geometric TTL coupling originates from the situation whereby one of the beams, usually the one carrying the main measurement signal, tilts and deviates from the initial interferometer light path, leading to a propagation distance greater than that of the completely aligned reference beam. The optical path difference between beams results in a phase difference, for example, the power fluctuation of the photodiode in a zero-difference interferometer, or the phase shift of the heterodyne signal in a heterodyne interferometer. 
\begin{figure}[ht!]
    \centering
    \includegraphics[width=0.7\linewidth]{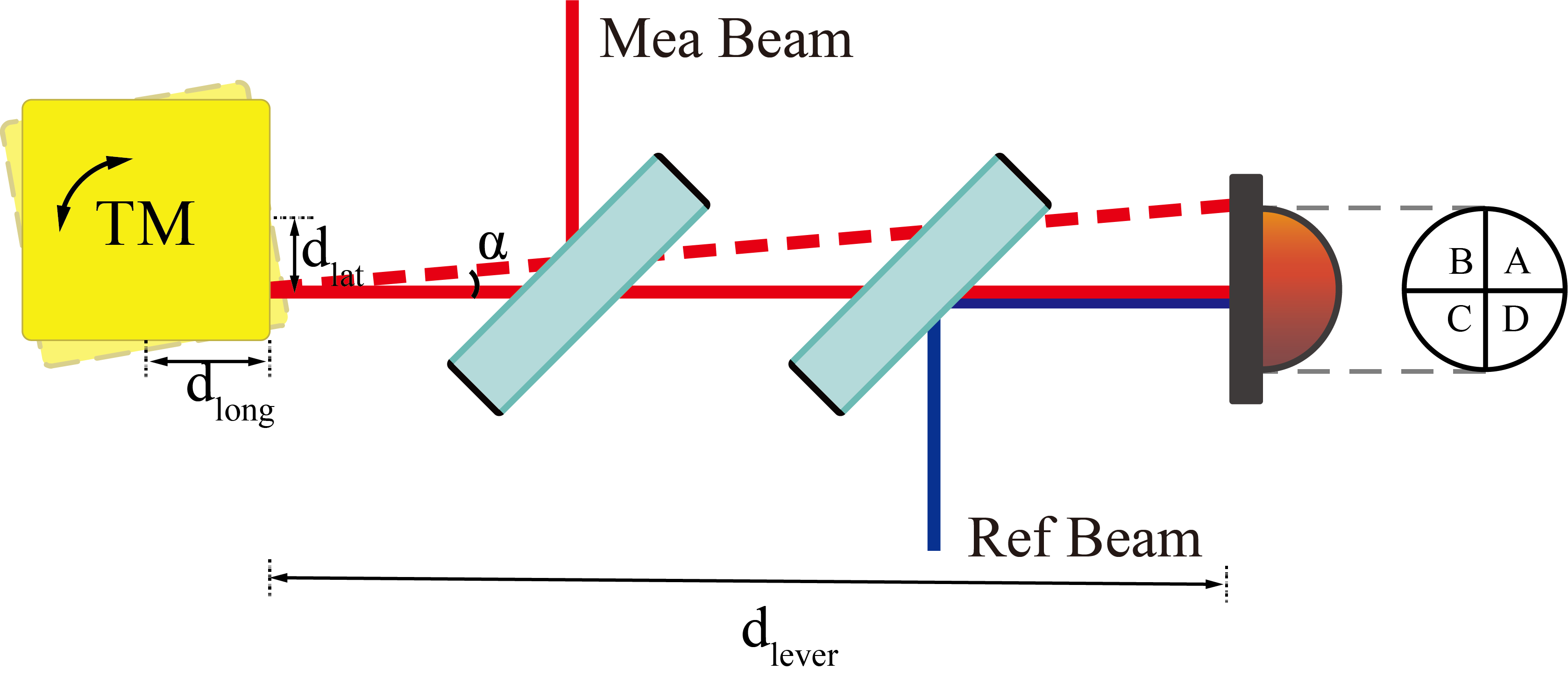}
    \caption{Geometric TTL coupling effects in TM interferometers: light lever effect and piston effect. When the test mass tilts (beam tilt angle $\alpha$), due to the presence of a lever arm $\mathrm{d_{lever}}$ between the TM surface and the detector, an extra optical path is generated, known as the light lever effect. The measurement light reflects on the TM surface, and its reflection point does not coincide with the center of mass (TM rotation center), leading to a longitudinal offset $\mathrm{d_{long}}$ and a lateral offset $\mathrm{d_{lat}}$. When the test mass tilts, an extra optical path is generated, known as the piston effect.}
    \label{fig1}
\end{figure}

Taking the TM interferometer as an example, when the TM tilts, geometric TTL coupling errors can occur due to the presence of a lever arm between the TM and the detector, a phenomenon known as the light lever effect. As shown in Fig. 1, assuming the laser is a plane wave, the expression for this effect can be calculated through simple geometric relationships as:
\begin{equation}
\mathrm{OPD}_{\text {lever }} \approx \frac{d_{\text {lever }}}{2}\alpha^2.
\end{equation}
Furthermore, an offset exists between the reflection point of the measuring light on the TM surface and the TM rotation point (center of mass), leading to an extra optical path, known as the piston effect. Its expression can be written as:
\begin{equation}
\mathrm{OPD}_{\text {piston }} \approx- d_{\text {lat }} \alpha+ d_{\text {long }} \alpha^2.
\end{equation}
where $\mathrm{d_{lever}}$ is the lever arm length, $\mathrm{d_{lat}}$ is the lateral offset of the reflection point, $\mathrm{d_{long}}$ is the longitudinal offset, and $\alpha$ is the beam tilt angle. All the above results are retained to the second order of the tilt angle $\alpha$.

However, in a real heterodyne interferometer, lasers have their own wavefront curvatures and power distributions. For instance, the TM interferometer uses Gaussian light while the interstellar interferometer uses both Gaussian light and flat-top light. If the two interference lasers are not perfectly matched and aligned on the detector surface (for example, due to laser axis offset, wavefront tilt, and distortion), they would interfere with the interference light field distribution on the detector surface and subsequently impact the final sensitive-axis optical path signal readout. This type of effect is known as a non-geometric TTL coupling. The sensitive axis optical path signal LPS readout can be written as the phase of the interference light field at the active area of the detector, integrated over that part,
\begin{equation}
\mathrm{LPS}=\frac{1}{k} \arg \left(\iint E_m E_r^* d S_{\mathrm{PD}}\right).
\end{equation}
Where $\mathrm{E_{m}}$ represents the electric field of the measurement light, and $\mathrm{E_{r}}$ refers to the electric field of the reference light, any factors affecting this integral will cause TTL coupling, such as the previously mentioned mismatch of light parameters, properties of the photodetector \cite{schuster2017tilt}, and even different definitions of the sensitive axis optical path signal can result in various TTL couplings, as mentioned in literature \cite{wanner2015brief}. In interstellar interferometers, the interfering beams are Gaussian light and flattop light, and the sensitive axis optical path signal needs to replace the electric field expression in the integral equation. The TTL coupling generation mechanism is roughly the same in both kinds of interferometers. However, due to the intercept and magnification of the far-end light by the interstellar telescope, the beam tilt angle is amplified, thereby affecting the level of TTL coupling.

\begin{figure}[ht!]
    \centering
    \includegraphics[width=0.85\linewidth]{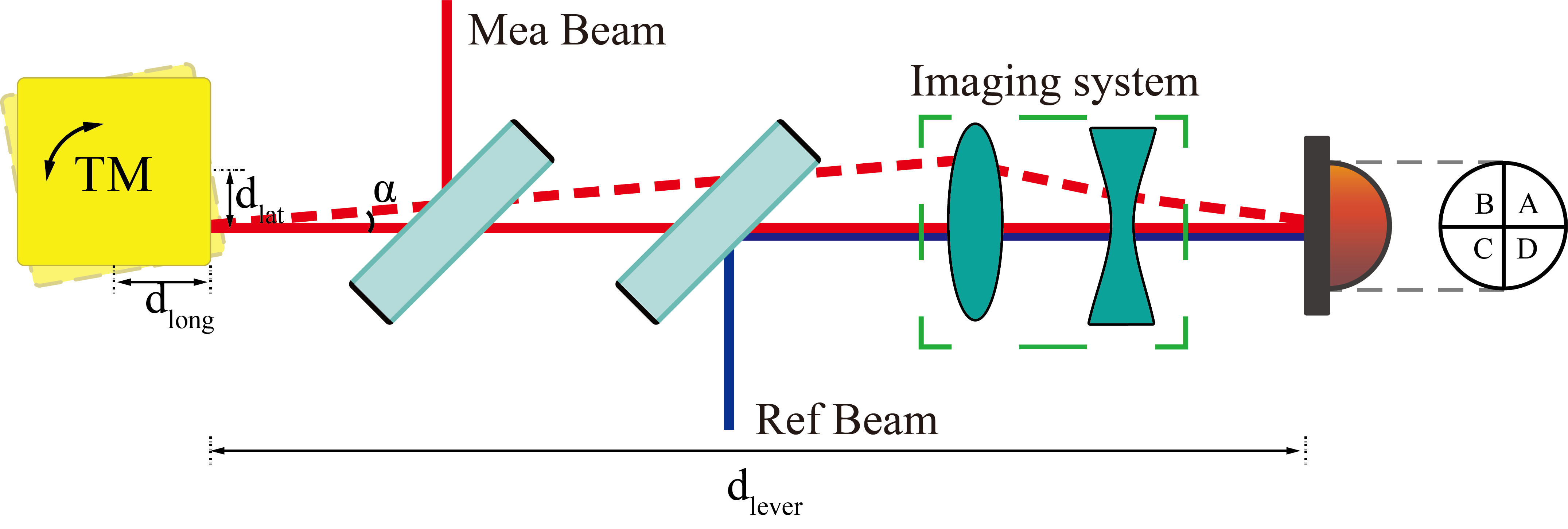}
    \caption{Suppression of TTL coupling in TM interferometers by imaging systems. By utilising a dual-lens imaging system, the optical path of the measurement light can remain unchanged, and lateral displacement of the light spot on the detector surface can be reduced, thus avoiding the generation of extra non-geometric TTL coupling effects. Furthermore, the residual TTL coupling can be compensated by deliberately introducing extra TTL coupling through adjustments in the position of the imaging system or the QPD.}
    \label{fig2}
\end{figure}

\subsection{Imaging system}
Imaging systems are an effective method for suppressing the TTL coupling \cite{Sonke2016,Tröbs2018,Chwalla2020}, as shown in Fig. 2. A strict optical design is employed to place the TM reflection surface in the entrance pupil of the imaging system and the quadrant photodiode (QPD) in the exit pupil. The Fermat principle is utilized to suppress the change in the geometric optical path of the measurement beam during tilting, as well as to restrain the offset of the light spot on the surface of the photodetector. However, imaging systems can alter the optical parameters of the interfering beams and the tilt angle of the measurement beam, potentially increasing non-geometric TTL coupling. Consequently, imaging systems can suppress geometric effects while leaving residual non-geometric effects. To reduce the system's TTL coupling below the required level, modifying the design of the imaging system is the only method to deliberately introduce optical path changes to offset non-geometric TTL coupling.

In this paper, we design a set of classic confocal imaging systems using existing lab lenses. Through fine-tuning the installation positions of the imaging system and QPD, we attempt to combine and test their suppression effects on TTL coupling within the system. The imaging system is installed on a three-axis displacement stage, allowing for precise multiaxis adjustments of the lens position. The detector installation is based on thermally stable single-blade aluminum flexors with ultrafine precision screws, achieving micrometer-level positional accuracy and milliradian angular accuracy.

\section{APTA: Advanced Pure Tilt Actuator}
\subsection{Measurement principle}
\begin{figure}[ht!]
    \centering
    \includegraphics[width=0.7\linewidth]{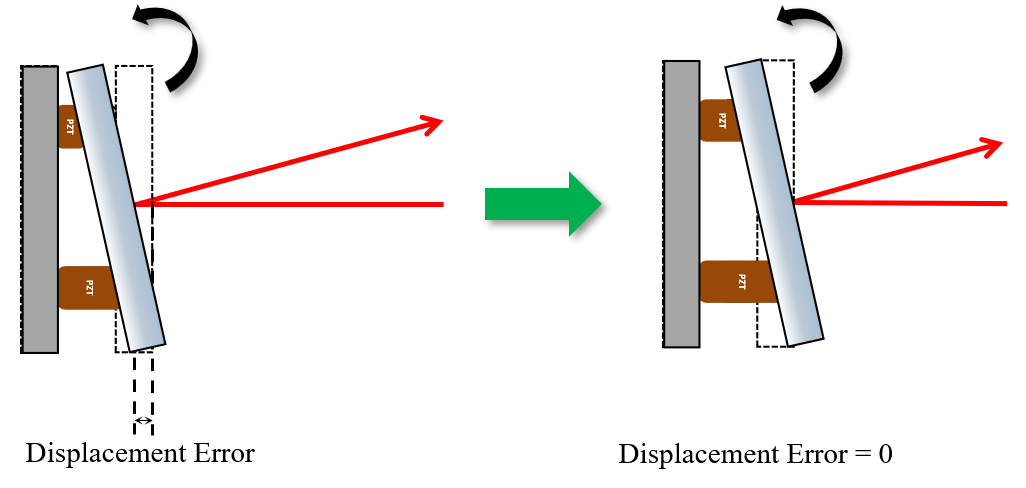}
    \caption{Residual longitudinal displacement generated by the tilt actuator during its motion. When the rotation point does not coincide with the reflection point of the measuring light, additional longitudinal displacement errors will occur.}
    \label{fig3}
\end{figure}
In TTL coupling calibration, it is necessary to subtract the residual longitudinal displacement of the tilting actuator at the measurement light reflection point. This requires monitoring the three degrees of freedom (longitudinal displacement, yaw and pitch) at the reflection point to effectively cancel out the actuator's residual noise. As I understand it, two non-collinear lines define a plane, and similarly, three non-collinear points determine a plane. During plane movement, we can infer the three degrees of freedom at any point on the plane from the displacement information at these three points. By leveraging this property, we can monitor and subtract the residual longitudinal displacement at the measurement light reflection point while also reading the tilt angle of the measurement light. To achieve these objectives, two critical conditions must be met. Firstly, the precise measurement of the longitudinal displacement information at three points on the plane; Secendly, Accurate positioning of these three points.
\begin{figure}[ht!]
    \centering
    \includegraphics[width=0.5\linewidth]{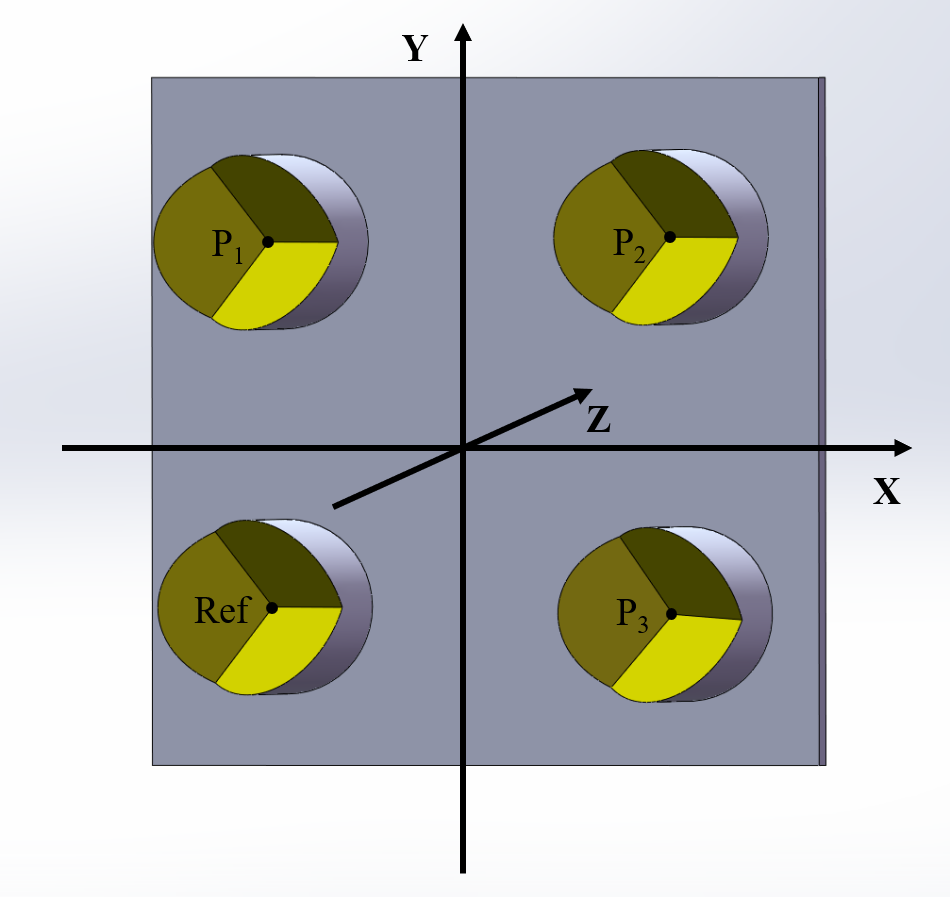}
    \caption{Diagram illustrating the plane of the pyramids and its installation position. Four pyramids are mounted on the base, with their vertices labeled clockwise as P1, P2, P3 and Ref. The last one is a spare pyramid with no laser incident on it.}
    \label{fig4}
\end{figure}

Once these conditions are satisfied, we can theoretically decouple the three degrees of freedom at the measurement light reflection point. As shown in Figure 4, assume three points on the same plane in space have coordinates $\mathrm{P}_1\left(\mathrm{x}_1, \mathrm{y}_1, \mathrm{z}_1\right), \mathrm{P}_2\left(\mathrm{x}_2, \mathrm{y}_2, \mathrm{z}_2\right) \text { and } \mathrm{P}_3\left(\mathrm{x}_3, \mathrm{y}_3, \mathrm{z}_4\right)$, where the Z-axis is the sensitive direction. Given that the plane's rotation angle is very small (of the order of a hundred microradians), we assume the coordinates remain unchanged, treating $\mathrm{x}_{1,2,3}$ and $\mathrm{y}_{1,2,3}$ as constants during the tilt, while the longitudinal displacements $\mathrm{z}_{1,2,3}$ change over time. The normal vector of the plane containing the mirror can be expressed through the cross-product of vectors between the three points:
\begin{equation}
\vec{n}=\overrightarrow{P_1 P_2} \times \overrightarrow{P_2 P_3}=\left(n_x, n_y, n_z\right).
\end{equation}
Substituting the coordinate information, the elements of the normal vector can be expressed as:
\begin{equation}
\begin{aligned}
& n_x=z_1 \cdot\left(y_3-y_2\right)+z_2 \cdot\left(y_1-y_3\right)+z_3 \cdot\left(y_2-y_1\right), \\
& n_y=z_1 \cdot\left(x_2-x_3\right)+z_2 \cdot\left(x_3-x_1\right)+z_3 \cdot\left(x_1-x_2\right), \\
& n_z=x_1 \cdot\left(y_2-y_3\right)+z_2 \cdot\left(y_3-y_1\right)+z_3 \cdot\left(y_1-y_2\right).
\end{aligned}
\end{equation}
where $n_{z}$ is a constant. Any point $P(x,y,z)$ on the plane containing $P_{1}$,\,$P_{2}$\,and $P_{3}$ can be expressed as:
\begin{equation}
z=-\frac{n_x \cdot\left(x-x_1\right)+n_y \cdot\left(y-y_1\right)}{n_z}+z_1.
\end{equation}

As can be seen from the equation above, knowing the longitudinal displacement and position information of three points on the plane enables accurate measurement of the longitudinal displacement z at any other point on the plane. By utilizing this property, the longitudinal displacement of the measurement light reflection point can be measured and the actuator's residual noise can be subtracted, thus calibrating the TTL coupling error in the interferometer more accurately.

\subsection{Optical design}
The APTA (Advanced Pure Tilt Actuator) consists of two parts: a compact four-beam interferometer and a conical rotary actuator, as shown in Figure 5. The compact four-beam interferometer is composed of a collimator, a lateral displacement prism, a beam-splitting prism, and a punctured plane mirror, with all optical elements adhered to an ultra-low expansion microcrystalline glass substrate using UV glue \cite{lin2023construction}, ensuring a structure with robust thermal stability.
\begin{figure}[ht!]
    \centering
    \includegraphics[width=0.8\linewidth]{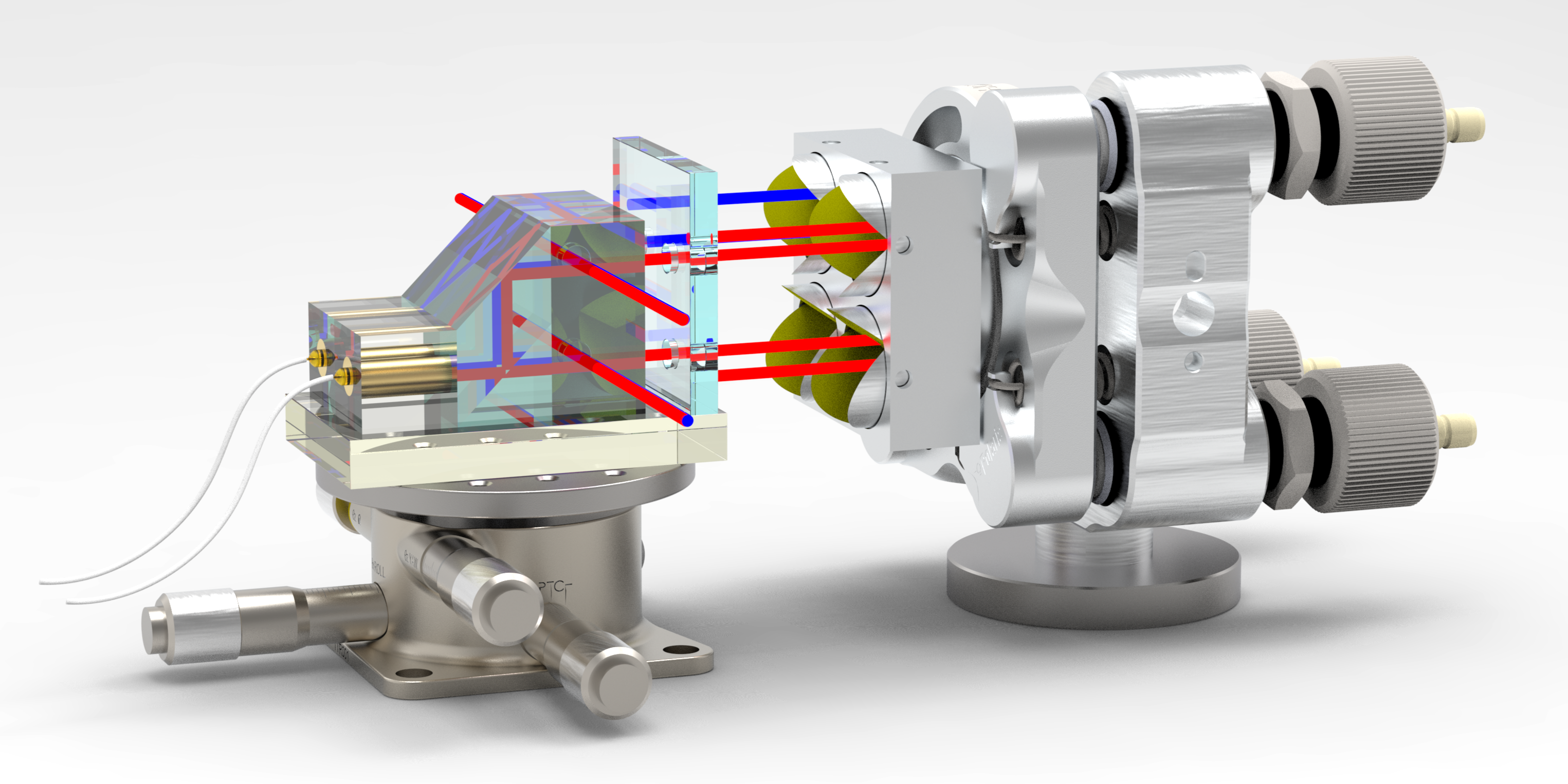}
    \caption{3D Structural Diagram of APTA. APTA consists of a compact four-beam interferometer and a corner-cube retroreflector. The interferometer is bonded using UV glue onto a substrate made of ultra-low expansion Zerodur, ensuring excellent thermal stability and a compact overall structure. APTA features three measurement interferometers and one reference interferometer. By measuring the movement of the target corner-cube, it decouples the three degrees of freedom motion at any point on the reflection plane. This system can be utilized for subtracting parasitic displacement noise from the actuator in TTL coupling experiments. Furthermore, the combination of a perforated mirror and the corner-cube prevents lateral shifts of the measurement beam during tilt and maintains high interference contrast throughout the process.}
    \label{fig5}
\end{figure}

The optical path design is shown in Figure 6. The laser emitted from the collimator is split into two layers, upper and lower, with four beams in total, by the lateral displacement prism. The two layers of beams are completely parallel. Each beam is then split again via the beam-splitting prism, with three of the beams exiting after being transmitted through the punctured reflecting mirror. The position of the hole in the mirror is precisely measured to ensure that there is no truncation of the light spot. The remaining transmitted laser beam reflects off the mirror and returns along the original optical path, interfering with another beam reflected from the beam-splitting prism. Since the reflecting mirror does not move, this interference signal serves as a reference to remove noise from the upstream fiber link and environmental noise. The conical rotary actuator consists of a piezoelectric ceramic mirror flipper and three hollow conicals. Piezoelectric ceramic acts as an actuator by providing three degrees of freedom (pitch rotation, yaw rotation, and longitudinal translation) controlled by voltage. The hollow conicals serve as target mirrors, reflecting the three transmitted beams back to the punctured reflecting mirror. Thanks to the property of conical prisms, which keeps the incident and outgoing beams parallel, adjusting the attitude of the perforated reflector will ensure each beam reflected by the cone exits the same way it entered. Then these interfere with their corresponding reference light, forming three independent measurement signals.
\begin{figure}[ht!]
    \centering
    \includegraphics[width=1\linewidth]{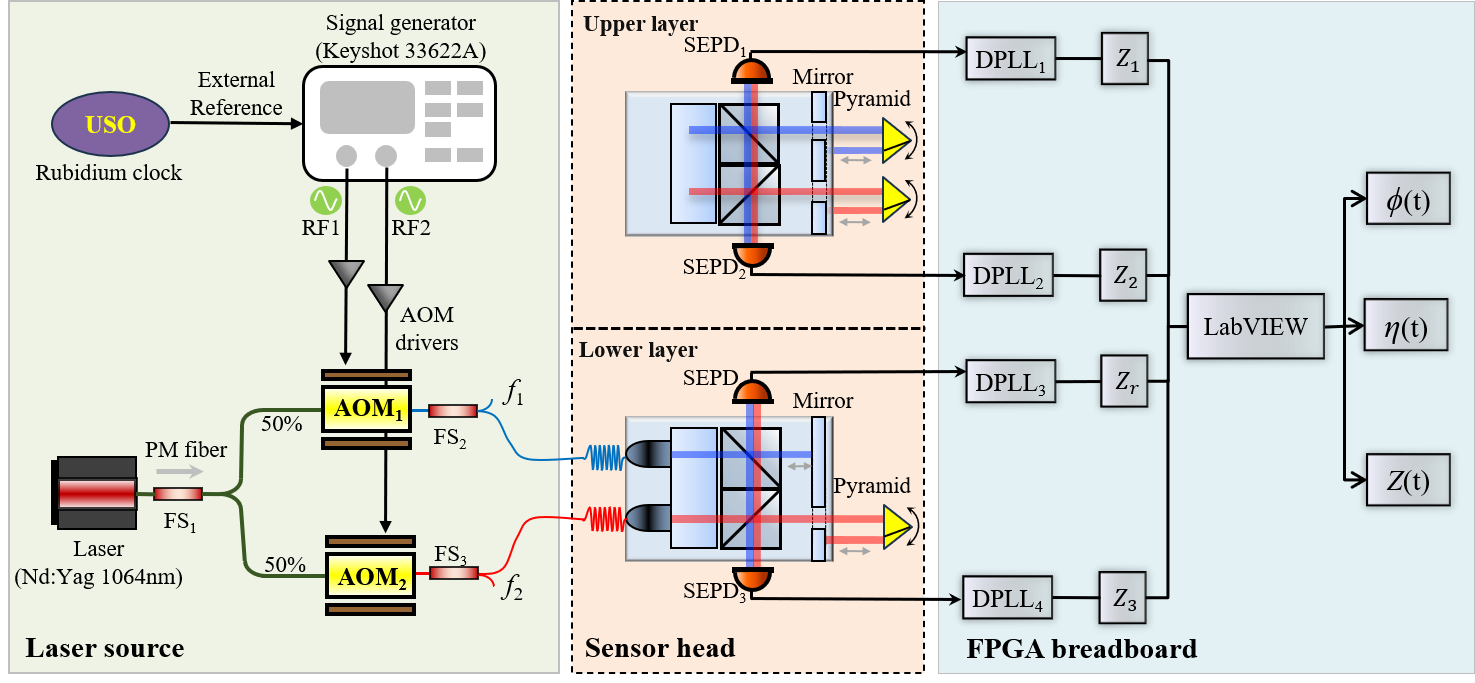}
    \caption{Optical Path Diagram of APTA. The optical path of APTA consists of two strictly parallel layers, each containing two interferometers, resulting in a total of four interferometers. Among them, three are measurement interferometers. The measurement light passes through a beam splitter and a perforated mirror, reaching the target corner-cube. The measurement beam, after passing through the corner-cube, reflects off the surface of the perforated mirror. By adjusting the orientation of the perforated mirror, the measurement beam retraces its path, retroreflects from the corner-cube, and returns along the original optical path to combine with the reference beam, generating the measurement interference signal. Additionally, in the reference interferometer, the measurement beam is directly reflected by the perforated mirror back along the original path to interfere with the reference beam, forming the reference signal. This reference signal is used to subtract the frontend fiber noise and environmental noise from the system.}
    \label{fig6}
\end{figure}

APTA's unique structure provides several advantages. First, when the target cone rotates, the contrast of the interference signals hardly changes, maintaining a high level. This is because the measurement beam always returns along the same path, thus ensuring consistently well-aligned interfering light \cite{lin2024compact}. Second, the optical path within the hollow cones is a constant value, solely dependent on the height of the cones and not the angle of incidence. This makes the cone insensitive to the angle of incidence - TTL coupling can thus be considered insensitive to an interferometer using a cone as the target mirror. Third, its versatile use. As an independently designed and built experimental platform, the overall structure of APTA is compact and easy to transport. Therefore, it can be used not only for specific TTL coupling experiments but also for other experiments requiring multi-degree of freedom motion. Fourth, APTA has three independent measurement interferometers that can detect the three degrees of freedom motion information of the target mirror. Moreover, multi-beam measurement has a larger linear range and lower non-linearity [20], allowing for future comparison and calibration of the DWS signal in the test platform.

\subsection{Error analysis}
APTA will be used for TTL coupling calibration, so it is necessary to analyze its own errors to avoid introducing them into the calibration results. The first concern is the dihedral angle error \cite{wei2010}. Due to the machining process, it is impossible to ensure that the three faces of the corner cube prism are perfectly perpendicular to each other, which affects the property of the constant optical path within the corner cube, thereby introducing calibration errors into the APTA. In this context, we need to account for the dihedral angle errors in the internal optical path calculations of the corner cube. Let's consider the dihedral angle errors as $\delta_{12}$ and $\delta_{13}$, where $\delta_{12}$ is the dihedral angle error between the first and second reflective surfaces, and $\delta_{13}$ is the dihedral angle error between the first and third reflective surfaces.

\begin{figure}[ht!]
    \centering
    \begin{subfigure}{0.4\textwidth}
        \centering
        \includegraphics[width=\textwidth]{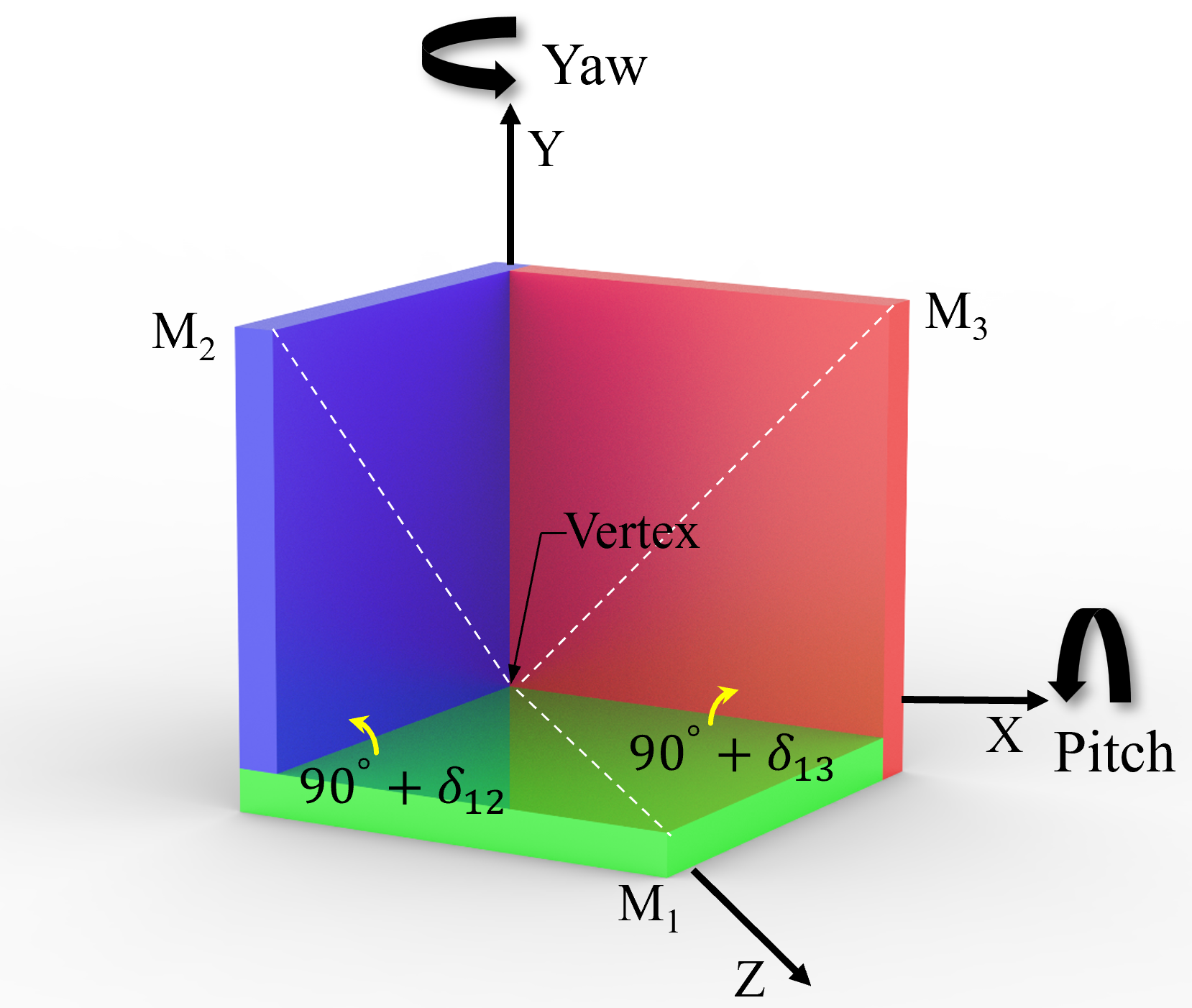}
        \caption{Dihedral angle errors}
        \label{fig:sub1}
    \end{subfigure}
    \hfill
    \begin{subfigure}{0.55\textwidth}
        \centering
        \includegraphics[width=\textwidth]{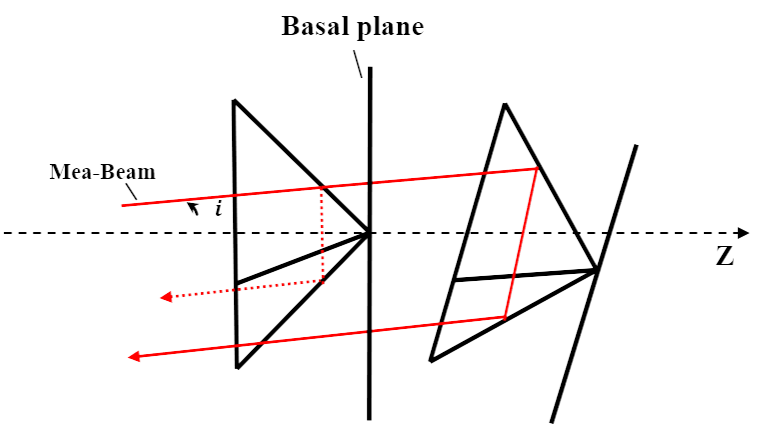}
        \caption{incident angle error}
        \label{fig:sub2}
    \end{subfigure}
    \caption{(a)The geometric structure of the pyramid and its position. Among them, $\mathrm{M}_{1}$, $\mathrm{M}_{2}$ and $\mathrm{M}_{3}$ are the three faces of the corner reflector, and their dihedral angle errors are $\delta_{12}$ and $\delta_{13}$. During the calculation, the vertex of the corner reflector is placed at the origin of the Cartesian coordinate system. (b)Additional optical path errors caused by the incidence angle when light is incident on the APTA plane at a non-perpendicular angle. $i$ is the angle of incidence of the measuring light relative to the plane of the pyramid.}
    \label{fig7}
\end{figure}
At this point, we only consider the change in the geometric optical path of the measurement light within the corner cube. Assuming the measurement light is a ray, we first assume that the vertex of the corner cube is positioned at the origin "$(0,0,0)^{\mathrm{T}}$"  of the coordinate system. The installation position of the corner cube is as shown in Figure 7(a). The normal vector of the exit plane is $\overrightarrow{n_0}=(0,0,1)^T$, and this plane is 1 meter away from the origin. The exit point of the light ray is $\overrightarrow{P_0}\left(x_0, 0,0\right)$, with the propagation direction $\vec{l}=(0,0,-1)^T$. Using ray-tracing methods for analysis and calculation, we can derive the analytical solution for the direction of the exit ray after reflection by the corner cube with errors (accurate to the second order of the dihedral angle error),
\begin{equation}
\overrightarrow{l^{\prime}} \approx\left(\begin{array}{l}
0 \\
0 \\
1
\end{array}\right)+\frac{1}{2}\left(\begin{array}{c}
-2 \delta_{12}-2 \sqrt{2} \delta_{13}-4 \delta_{12} \delta_{13}+\sqrt{2} \delta_{12}^2 \\
-2 \sqrt{2} \delta_{12}+2 \sqrt{2} \delta_{12} \delta_{13}-\delta_{12}^2+2 \delta_{13}^2 \\
-3 \delta_{12}^2-2 \sqrt{2} \delta_{12} \delta_{13}-2 \delta_{13}^2
\end{array}\right)+\left(\begin{array}{c}
-\delta_{12}+\sqrt{2} \delta_{13} \\
-\delta_{12}+\sqrt{2} \delta_{13} \\
0
\end{array}\right)\left(\begin{array}{l}
\eta \\
\phi \\
0
\end{array}\right).
\end{equation}
Here, $\eta$ and $\phi$ represent the pitch and yaw angles of the corner cube prism, respectively. The first term in the equation can be understood as the direction of the reflected ray after passing through an ideal corner cube reflector, which is completely parallel and opposite to the direction of the incident ray. The second term represents the deviation in the direction of the exit ray when the corner cube reflector with dihedral angle errors is in a static state. The third term represents the coupling deviation in the direction of the exit ray due to the rotation of the corner cube reflector with dihedral angle errors. During the rotation of the corner cube reflector, the introduced geometric optical path error is:
\begin{equation}
\begin{aligned}
    \text{OPD}_{\text{error}} \approx & \underbrace{\left( x_0 \delta_{12} + \sqrt{2} x_0 \delta_{13} + \sqrt{2} \delta_{12} \delta_{13} + \left( \tfrac{3}{2} + \tfrac{x_0}{\sqrt{2}} \right) \delta_{12}^2 + \delta_{13}^2 \right)}_{\text{static term}} \\
    & + \underbrace{\left[ x_0 \delta_{12} - \sqrt{2} x_0 \delta_{13} \right] \eta}_{\text{first-order term}}+ \underbrace{\left\{ \left( 1 - \frac{x_0}{\sqrt{2}} \right) \delta_{12}^2 - 2 \delta_{13}^2 \right\} \eta}_{\text{first-order term}} \\
    &+ \underbrace{\left[ \left( \sqrt{2} - \frac{x_0}{2} \right) \delta_{12}^2 - \left( 2 + \sqrt{2} x_0 \right) \delta_{12} \delta_{13} + x_0 \delta_{13}^2 \right] \phi}_{\text{first-order term}} \\
    & + \underbrace{ \left[ -\frac{x_0}{2} \delta_{12} - \frac{x_0}{\sqrt{2}} \delta_{13} \right] \eta^2}_{\text{second-order term}}+ \underbrace{ \left[ -\sqrt{2} x_0 \delta_{12} \right] \eta \phi}_{\text{second-order term}}
\end{aligned}
\end{equation}
The first term in the equation (static term) represents the additional optical path introduced when the corner cube reflector is static, accounting for the exit point position and its dihedral angle error. The second and third terms represent the linear coupling of the pitch angle $\eta$. However, the third term involves second-order polynomials of the dihedral angle error, indicating that if the dihedral angle deviation is small, its contribution to the optical path length change is negligible. The fourth term denotes the change in optical path length corresponding to the second-order polynomial of the pitch angle $\eta$. Similarly to the third term, its impact is negligible for small rotation angles and dihedral angle errors. The fifth term shows the linear coupling of the yaw angle $\phi$. Given that the second-order polynomial of the dihedral angle deviation is very small, the influence of the fifth term on the optical path length change is not significant. Similarly, the contribution from the sixth term may be negligible.
\begin{figure}[ht!]
    \centering
    \begin{subfigure}{0.48\textwidth}
        \centering
        \includegraphics[width=\textwidth]{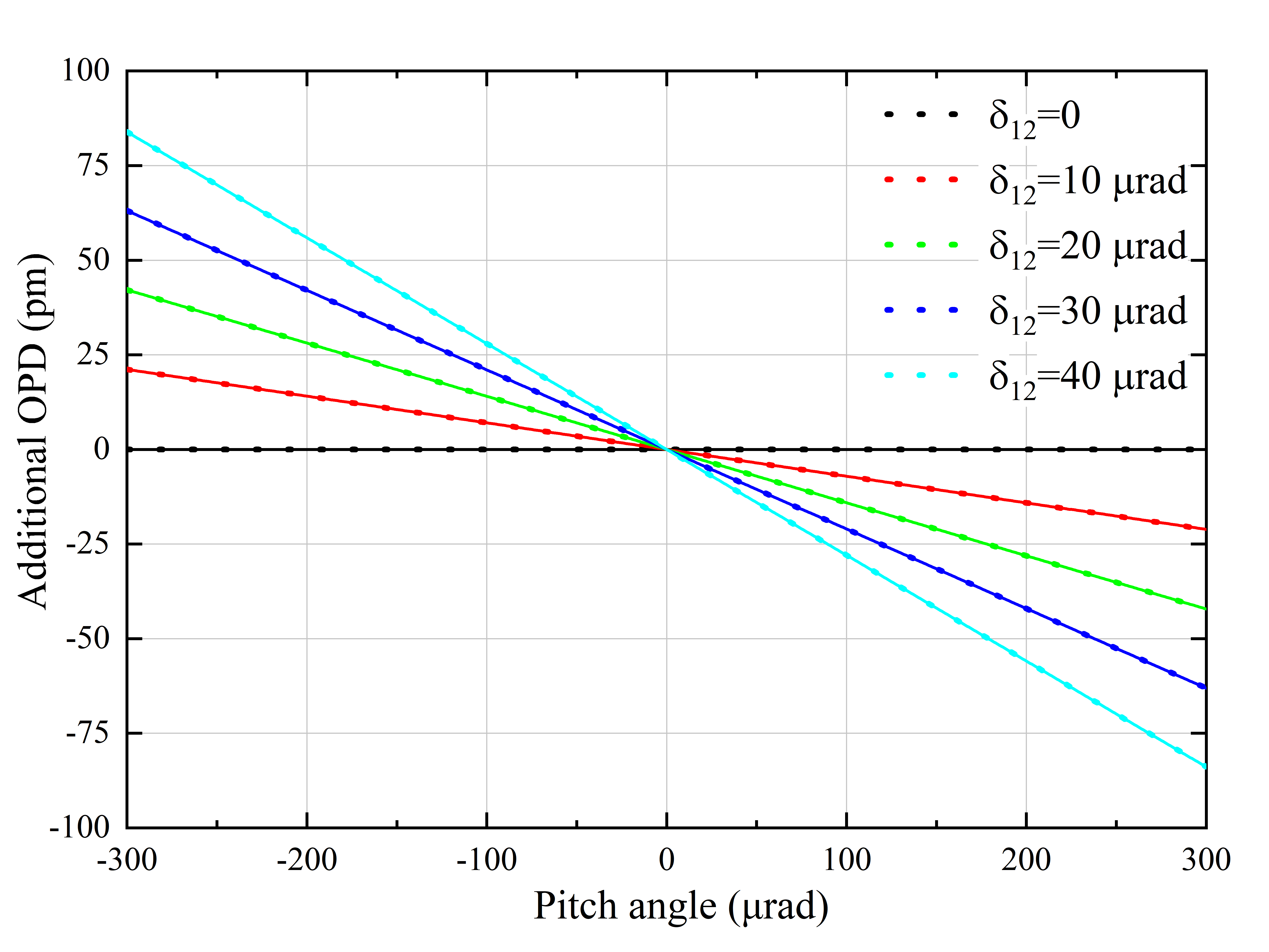}
        \caption{Dihedral angle errors}
        \label{fig8-1}
    \end{subfigure}
    \hfill
    \begin{subfigure}{0.48\textwidth}
        \centering
        \includegraphics[width=\textwidth]{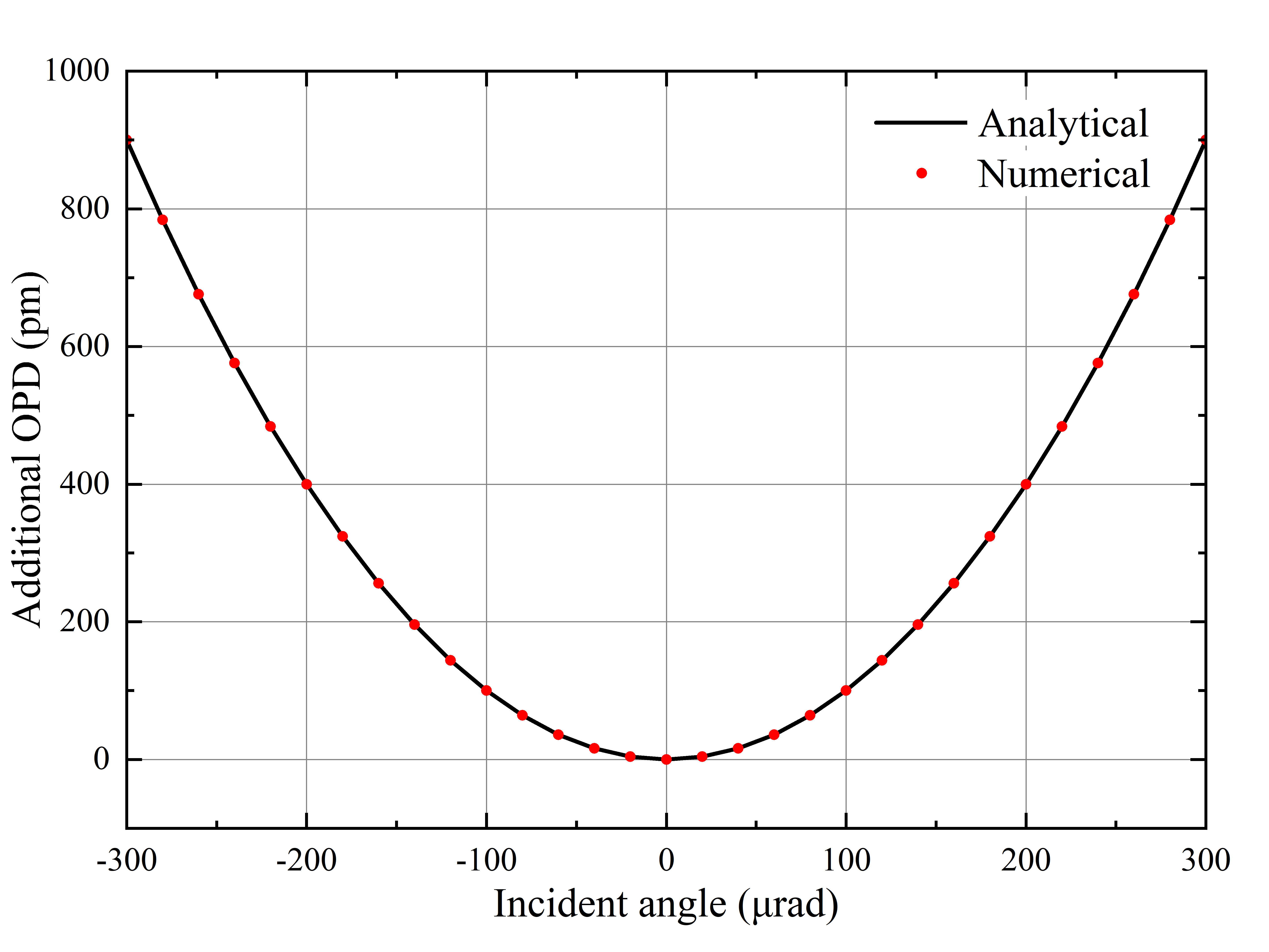}
        \caption{incident angle error}
        \label{fig8-2}
    \end{subfigure}
    \label{fig8}
    \caption{(a)The geometric optical path error caused by the dihedral angle error of the corner cube during pitch deflection. The additional OPD under five different dihedral angle errors was calculated. The additional OPD increases as the dihedral angle error increases. (b)Additional OPD induced by different incidence angles. The solid line represents the analytical calculation results, while the dashed line represents the numerical calculation results. The two sets of results overlap closely.}
    \label{fig:bothimages}
\end{figure}

In practice, the dihedral angle error can be controlled in 3 arcseconds ($~1.45 \mu \mathrm{rad}$), the deflection of the corner cube reflector is approximately $-300 \sim {300} \,\mu \mathrm{rad}$, and the accuracy of displacement measurement is within the picometer range, so the influence of the third-order term can be ignored. To verify the correctness of the analytical solution, we numerically solved the ray-tracing process using Matlab, and compared the results of the two methods. Assuming the exit point is $\vec{s}(0.005 \mathrm{~m}, 0,0)$ and the deflection of the corner cube reflector is $-300 \sim 300 \,\mu \mathrm{rad}$, and there is only one dihedral angle error $\delta_{12}=0$, $\delta_{13}=0\sim40\,\mu \mathrm{rad}$, only the coupling term (excluding the static term) is calculated.When the corner cube undergoes pitch deflection $\eta$, the additional optical path introduced by the coupling of different dihedral angle errors is shown in Figure 8(a), where the solid line represents the numerical calculation results and the dashed line represents the analytical calculation results. It can be seen that the maximum TTL coupling coefficient is $\sim0.28\,\mu \mathrm{m/rad}@\delta_{13}=0\sim40\,\mu\mathrm{rad}$, and the analytical results are in good agreement with the numerical results.

Next, we consider the incident angle error of the measurement light on the corner cube prism. Ideally, the three measurement laser beams should be parallel to each other in pairs and perpendicular to the plane of the corner cube; otherwise, the measurement interferometer will not be able to accurately measure the longitudinal displacement of the corner cube vertex. In the experiment, we first use a flat mirror to reflect the three measurement beams back along their original paths, while the other end of the mirror is adjusted using an independent interferometer to reflect its measurement beam back along its original path. During the experiment, the power of the light at the return port of the circulator can be used to determine whether the measurement beam is returning along its original path. When the return power of all measurement beams is maximized, the mirror is replaced with a corner cube combined with a mirror, which forms the corner cube rotation actuator. Then, by adjusting its alignment to restore the return power of all measurement beams, it can be assumed that the three measurement beams in the APTA are aligned: parallel to each other in pairs and perpendicularly incident onto the plane containing the corner cube vertex. Despite efforts to align the beam angles, some residual misalignment may still exist. As shown in Figure 7(b), assuming the incidence angle of the beam is $i=\delta$, the optical path $S_{\mathrm{pyramid}}$ of the laser inside the corner cube is calculated using the ray tracing method as follows,

\begin{equation}
S_{\text {pyramid }}=\frac{2 n \mathrm{H}}{\sqrt{1-\frac{\sin i^2}{n^2}}}.
\end{equation}
where $n$ is the refractive index of the corner cube material. For a hollow corner cube n, and H is the height of the corner cube (distance from the vertex to the incident plane). Since a hollow corner cube is used in this experiment, when the laser is incident non-orthogonally ($i \neq 0^{\circ}$) on the surface of the corner cube, the optical path error approximated to the second order is,
\begin{equation}
\mathrm{OPD}_{\text {error }} \approx 2 \mathrm{H} \cdot\left(1+\frac{i^2}{2}\right)-2 \mathrm{H}=\mathrm{H} i^2.
\end{equation}

In this experiment, the height of the corner cube $H$ is 0.01 m. From the above equation, the optical path difference under various incidence angles can be analyzed and compared with the numerical calculation results, as shown in Figure 8(b). The simulation results indicate that the optical path error increases with the incidence angle; therefore, the incidence angle should be kept as small as possible. Based on the current experimental testing level, the incidence angle should be controlled below $100\, \mu \mathrm{rad}$.

Finally, considering the installation errors related to the position of the corner cube, as derived in Section 3.1, it is necessary to combine the position and displacement information of the corner cube vertex to obtain the displacement information at the measurement light reflection point in the TTL coupling experiment. Therefore, the position of the pymarid vertex is crucial for the final calibration of the TTL coupling. In this experiment, a perforated metal base is used to install the hollow corner cube prism, which is then fixed by metal-set screws. This means that the installation accuracy of the corner cube vertex position depends on the machining accuracy. Ideally, the coordinates of the vertices of three corner cubes in the XY plane are $P_1(-7 \mathrm{~mm}, 7 \mathrm{~mm}), P_2(7 \mathrm{~mm}, 7 \mathrm{~mm})$ and $P_3(7 \mathrm{~mm},-7 \mathrm{~mm})$. At present, the machining accuracy of the metal base is ±0.05 mm. Since the machining errors are random, in order to realistically reflect the actual situation, random errors within the range of ±0.05 mm  are assigned to the coordinates of the vertices of the three corner cubes in the XY plane, as shown in Table 1.

\begin{table}[h]
\centering
\caption{Initial Position and Errors of the Pyramids' Vertex}
\begin{tabular}{>{\centering\arraybackslash}m{1.5cm} >{\centering\arraybackslash}m{1cm} >{\centering\arraybackslash}m{1.5cm} >{\centering\arraybackslash}m{1.5cm} >{\centering\arraybackslash}m{1.5cm} >{\centering\arraybackslash}m{1.5cm}}
    \toprule
    \textbf{} & \textbf{Unit} & \boldmath{$x$} & \boldmath{$y$} & \boldmath{$\Delta x$} & \boldmath{$\Delta y$} \\
    \midrule
    Pyramid1 & mm & -7 & 7 & 0.0477 & -0.0216 \\
    Pyramid2 & mm & 7 & 7 & -0.0325 & 0.0189 \\
    Pyramid3 & mm & 7 & -7 & 0.0235 & 0.0481 \\
    \bottomrule
\end{tabular}
\end{table}

\begin{figure}[ht!]
    \centering
    \includegraphics[width=0.6\linewidth]{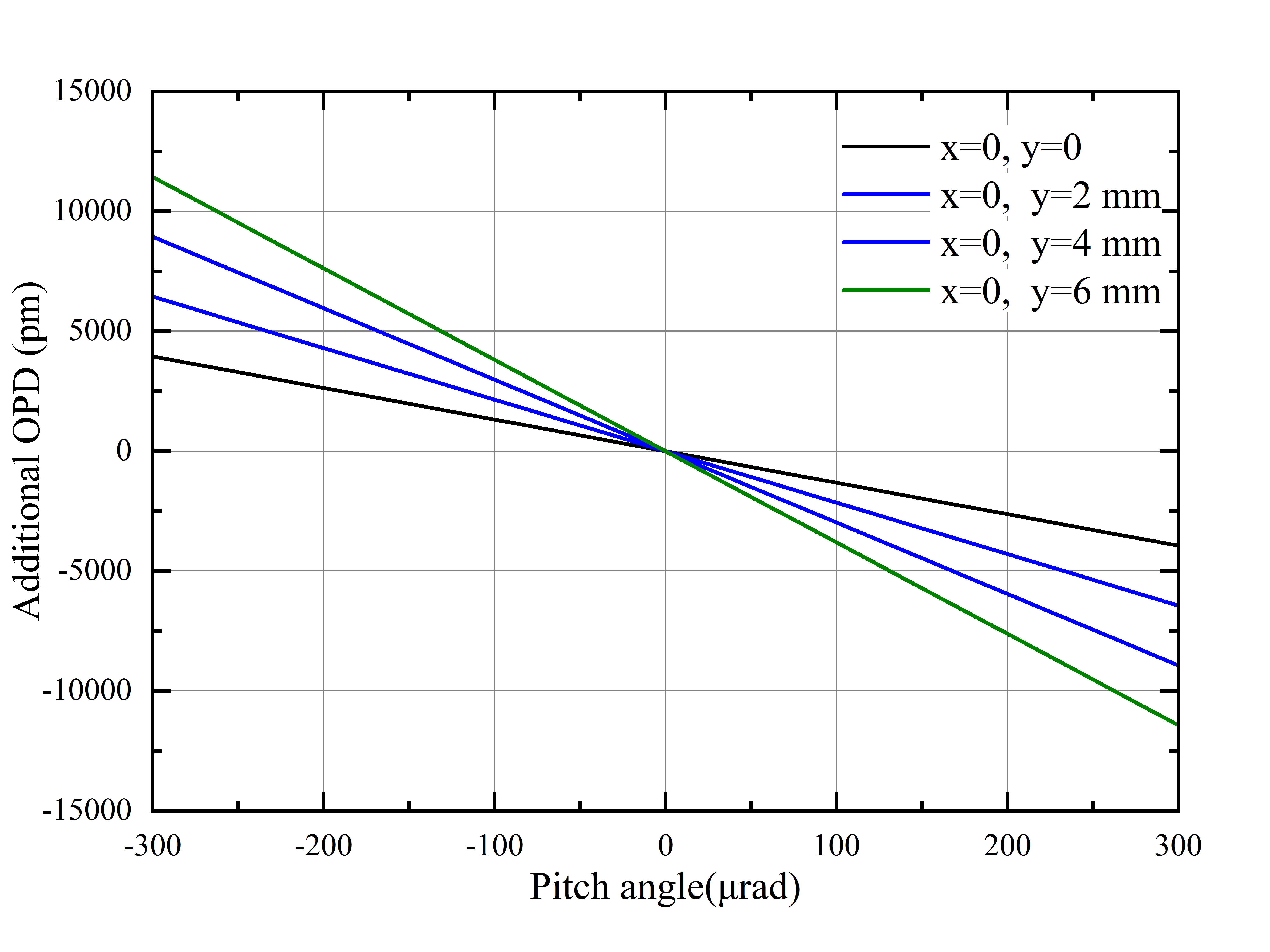}
    \caption{The additional OPD caused by the installation error of the corner reflector position shows a clear linear correlation.}
    \label{fig9}
\end{figure}

Assuming only pitch rotation around the X-axis occurs, substituting the above errors into formulas (6, 7), the errors in the three degrees of freedom at the origin O are given by:
\begin{equation}
\begin{aligned}
& \Delta \phi_{\text {error }} \approx 5.831 \times 10^{-3} \cdot \eta \\
& \Delta \eta_{\text {error }} \approx-4.157 \times 10^{-3} \cdot \eta \\
& \Delta z_{\text {error }} \approx 1.317 \times 10^{-3} \cdot \eta
\end{aligned}
\end{equation}
where $\phi$ is the yaw angle, $\eta$ is the pitch angle, and $\Delta\phi_{error}$, $\Delta\eta_{\mathrm{error}}$ and $\Delta z_{\mathrm{error}}$ are the yaw angle error, pitch angle error, and longitudinal displacement error at the origin, respectively. From the above equations, it can be seen that even if the plane of the corner cube only undergoes one degree of freedom motion, errors will still be generated in all three degrees of freedom. This means that the lateral displacement of the vertex position will lead to cross-coupling between each readout signal of the APTA. The longitudinal displacement error $\Delta z_{\mathrm{arb}}$ at any point on the vertex plane of the pymarid is expressed as:
\begin{equation}
\Delta z_{\text{arb}} \approx (-x_{\text{arb}} \cdot \Delta \phi_{\text{error}}) + (-y_{\text{arb}} \cdot \Delta \eta_{\text{error}}) + \Delta z_{\text{error}}.
\end{equation}
where $x_{\text{arb}}$ and $y_{\text{arb}}$ are the lateral position offsets of any point on the corner cube vertex plane with respect to the initial origin in the X and Y axes. Calculating the longitudinal displacement errors at three different positions, with coordinate information $x = 0$ mm,  $y = 0, 2, 4, 6$ mm, the results are shown in Figure 9. It can be seen that the longitudinal displacement increases with the increase in the offset, and exhibits a significant first-order linear relationship, which cannot be ignored.

To suppress the above errors, intentional introduction of errors can be used to offset the known errors. The specific method is to adjust the position of the measurement light-reflection point during the decoupling process to offset the existing linear errors. The feasibility of this method has been preliminarily verified in previous studies \cite{lee2021development}.

\section{Experimental setup and results}
\subsection{Experimental setup}
A ground-based experimental setup was established to test the TTL coupling in the quality inspection interferometer, where APTA and the imaging system are used respectively for calibration and suppression of the TTL coupling. As shown in Figure 10, the entire setup consists of two parts: the TM interferometer and APTA. The heterodyne interferometer is used to simulate the quality inspection interferometer, while the APTA is used to simulate the tilt of the TM attitude and to compensate for the errors of longitudinal displacement during motion. The following sections will explain each part in detail.
\begin{figure}[ht!]
    \centering
    \includegraphics[width=1\linewidth]{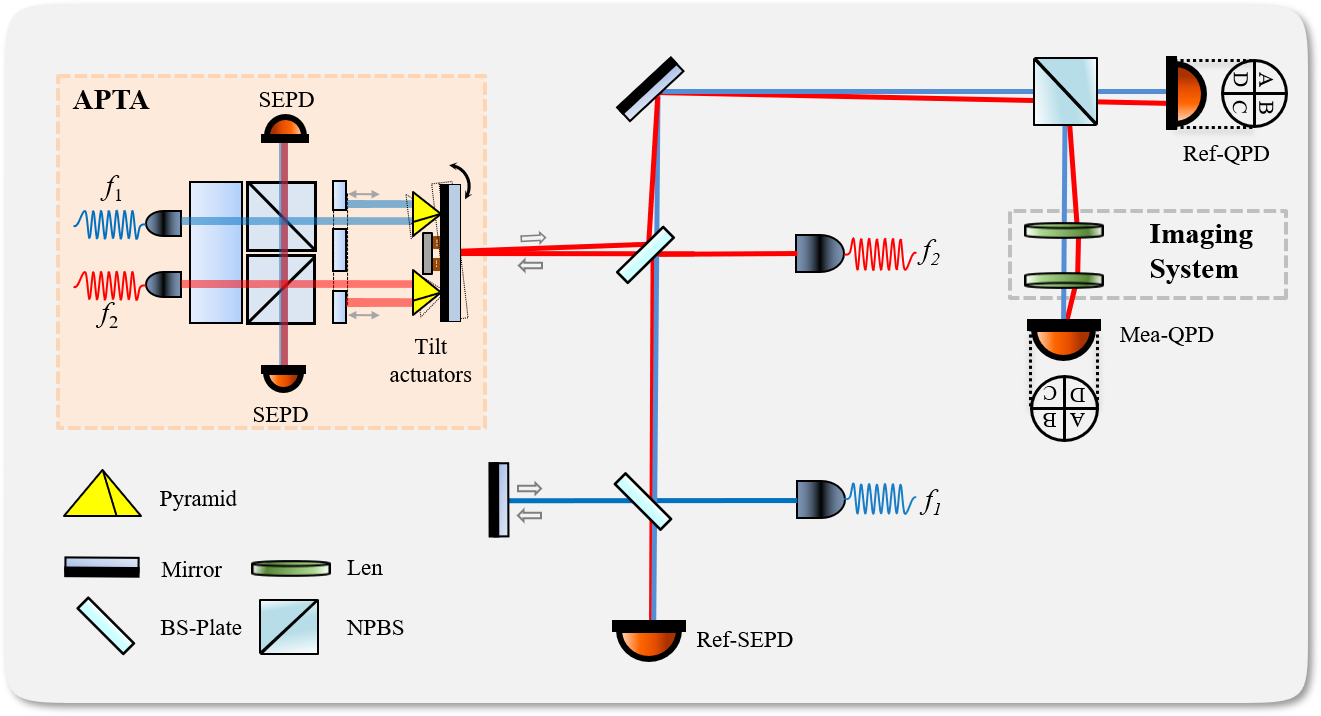}
    \caption{Schematic of the experimental setup.}
    \label{fig10}
\end{figure}

In this experiment, a low-noise narrow-linewidth NPRO laser (Coherent, Mephisto) with a central wavelength of 1064 nm was used as the light source. A high resolution dual channel waveform generator (Keysight, 33600A series) was utilized to generate two sinusoidal signals with frequencies $f_1 = 40$ MHz and $f_2 = 40.01$ MHz, respectively, achieving a heterodyne frequency of 10 kHz for the heterodyne interferometer. The orthogonal interference signals were received by a variable-gain InGaAs detector (Thorlabs, PDA20CS2) and a quadrant photodetector (Kyphotonics). Furthermore, a data acquisition (DAQ) system with low quantization loss (NI-PXIe 4464) was used for phase acquisition and demodulation. The sampling rate was set to 1 MHz, with a phase readout rate of 50 samples/s.

A single-frequency laser is transmitted to the optical workstation through a single-mode polarization-maintaining fiber. In the optical workstation, the laser undergoes frequency shifting via a pair of acousto-optic modulators. The two laser beams, which have a frequency difference of 10 kHz and nearly equal optical path length and power, are used as the measurement beam $f_1$ and the reference beam $f_1$, respectively, and are injected into the space interferometry portion. To transmit the two heterodyne beams into the APTA and the heterodyne interferometer, each beam is split using a 50:50 beam splitter, resulting in a total of four laser beams coupled into the fiber collimators.

In the TM interferometer, the measurement beam and the reference beam pass through nonpolarizing beam splitters, respectively, where the two reflected beams interfere. The interference beam is detected by a large-area single-quadrant photodetector through a mirror, forming a reference signal used to subtract fiber noise and environmental noise at the front end of the TM interferometer. The measurement beam, after passing through another nonpolarizing beam splitter and reflecting off the APTA mirror, generates beam tilt and reflects again through the. At this point, the tilted measurement beam and the reference beam are combined in NPBS, forming two separate interference beams. One of these interference beams is detected by a quadrant photodetector without passing through the imaging system, using differential wavefront sensing to obtain the tilt angle between the two beams and measure the TTL coupling magnitude within the TM. The other interference beam, passing through a confocal dual-lens imaging system, is detected by another quadrant photodetector. This setup is used to verify the imaging system's ability to suppress TTL coupling. The parameters for the imaging system and the TM interferometer are listed in Table 2.
\begin{table}[h]
\centering
\caption{Experimental setup parameters}
\begin{tabular}{>{\centering\arraybackslash}m{6cm} >{\centering\arraybackslash}m{2cm} >{\centering\arraybackslash}m{2cm}}
    \toprule
    \textbf{Parameter} & \textbf{Unit} & \textbf{Setting} \\
    \midrule
    distance from TM to Len1 & mm & $17 \,(f_1)$ \\
    distance from Len1 to Len2 & mm & $22 \,(f_1 + f_2)$ \\
    distance from Len2 to QPD & mm & $5 \,(f_2)$ \\
    Len1 focal length ($f_1$) & mm & $17$ \\
    Len2 focal length ($f_2$) & mm & $5$ \\
    Interference beam waist diameter & mm & $2.4$ \\
    \bottomrule
\end{tabular}
\end{table}

In high-precision experiments, interferometers are typically built on ultra-low expansion glass substrates and placed in a vacuum environment to suppress environmental disturbances such as temperature fluctuations and air currents. However, in this experiment, the test platform was constructed on a stainless steel honeycomb breadboard and placed in ambient air. The entire system was enclosed with an acrylic cover for passive temperature control and acoustic isolation. To further suppress noise that is not related to the tilt modulation signal and to more accurately calibrate the TTL coupling in the interferometer, we employed coherent filtering techniques \cite{Sonke2016}. This filtering technique requires applying a modulation signal to the tilt actuator and collecting multiple cycles of the interferometer signal. By performing a fast Fourier transform on the interferometer signal, we can consider any signal frequencies in the frequency domain that do not equal the modulation frequency or its harmonics as noise. By only computing the inverse transform corresponding to the modulation frequency and its harmonics, we retain the displacement signal related to the beam tilt and reconstruct the TTL coupling curve. To avoid frequency leakage, we applied a 0.5 Hz sinusoidal modulation signal to the tilt actuator, with the post-downsampled sampling rate of the phase meter set to 50 Hz, an integer multiple of the modulation frequency. Clock synchronization was performed between the modulation signal source and the phase meter, and the sampling time was approximately 10 minutes.
\subsection{Test results}
After completing this setup, we compared the interferometer signals on the reference QPD and the measurement QPD, which respectively represent the original TTL coupling in the TM interferometer and the suppressed TTL coupling after the imaging system. By comparing these two signals, we can assess the effectiveness of the imaging system in suppressing TTL coupling. In the experiment, the longitudinal displacement signal LPS was calculated using the arithmetic mean phase of the four quadrants of the QPD,
\begin{equation}
{\mathrm{LPS}}^{\text{AP}} = \frac{\phi_A + \phi_B + \phi_C + \phi_D}{4k}.
\end{equation}
where $\phi_A$ represents the phase measured by quadrant A of the QPD, and similarly for the other quadrants. $k = \frac{2\mathrm{\pi}}{\mathrm{\lambda}}$ is the wavenumber, with the laser wavelength being 1064 nm. The beam tilt angle information was obtained through the DWS signal.

The calibration results of TTL coupling in the TM interferometer before and after suppression by the imaging system are shown in Figure 11. After careful alignment and match of the reflection points \cite{lin2024compact}, the imaging system was aligned and the QPD measurement was adjusted to minimize TTL coupling. Without the imaging system, the longitudinal displacement readout of the system could change up to 3 nm within a tilt angle range of $\pm 200$ $\mu$rad. After incorporating the imaging system, the longitudinal displacement readout variation was reduced to below 0.6 nm. It is evident that the TTL coupling curve exhibits a significant quadratic effect with respect to angle, although higher-order terms still remain. To calculate the variation of the angle-to-twist ratio coefficient, we fitted the TTL coupling curve with a fourth-order polynomial and plotted the first derivative against the angle. It can be seen that within a beam tilt range of $\pm 100$ $\mu$rad, the TTL coupling coefficient, after suppression by the imaging system, decreased from $\pm 15$ $\mu$m/rad to below $\pm 5$ $\mu$m/rad.
\begin{figure}[ht!]
    \centering
    \begin{subfigure}{0.48\textwidth}
        \centering
        \includegraphics[width=\textwidth]{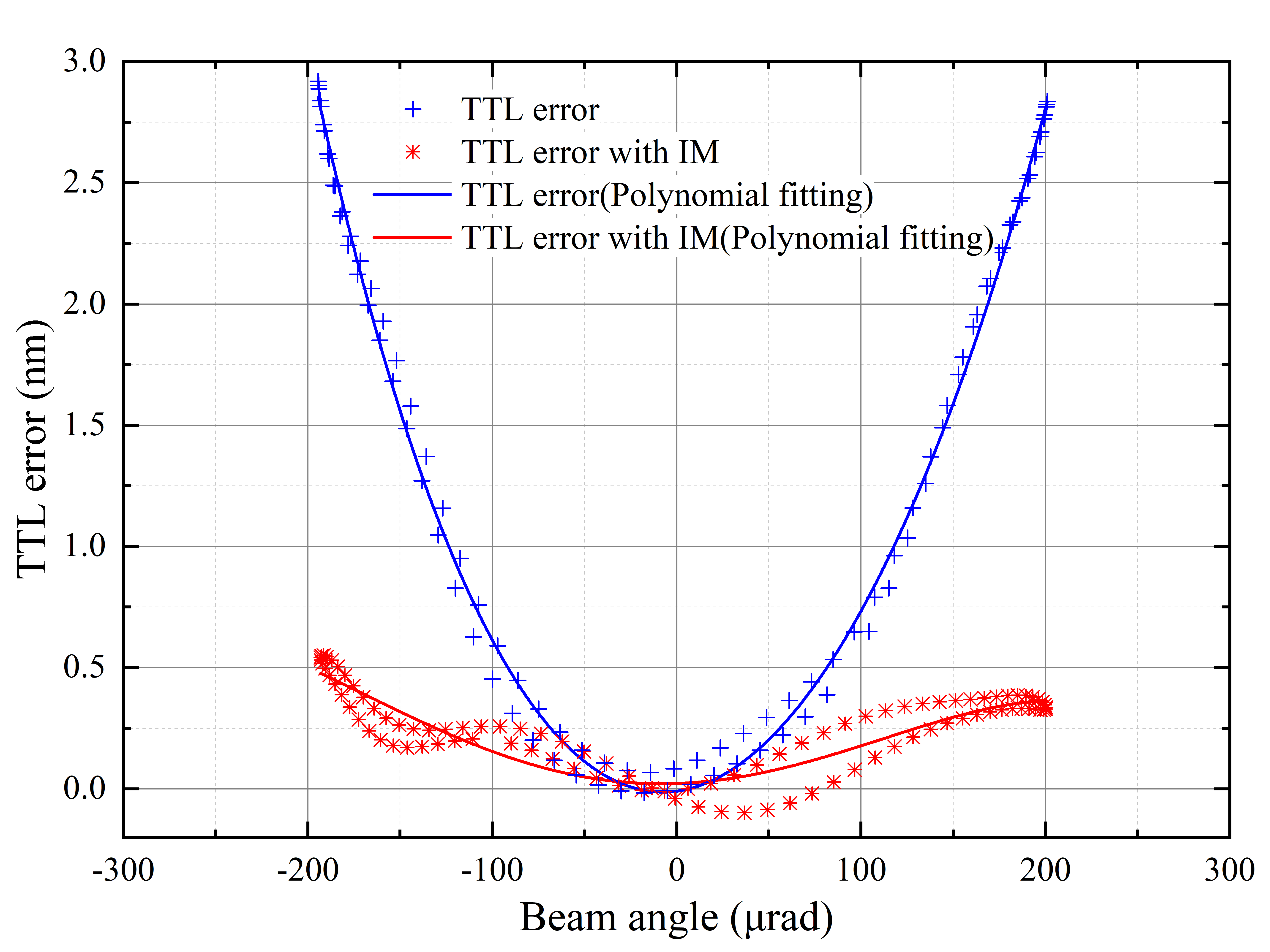}
        \label{fig11-1}
    \end{subfigure}
    \hfill
    \begin{subfigure}{0.48\textwidth}
        \centering
        \includegraphics[width=\textwidth]{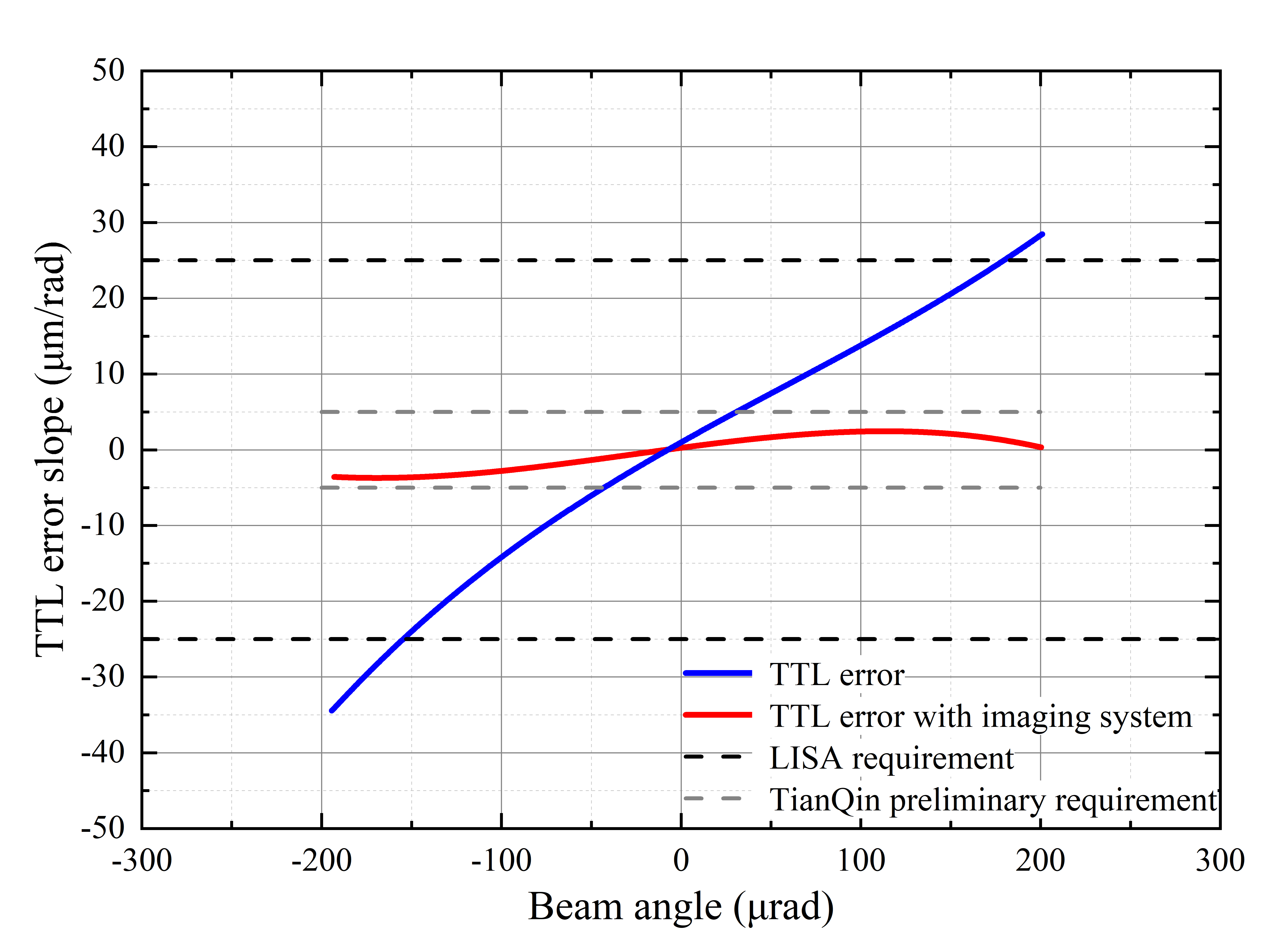}
        \label{fig11-2}
    \end{subfigure}
    \caption{TTL coupling in the TM interferometer and its coupling coefficient variation with angle before and after suppression by the imaging system. The dotted line represents the experimental data, and the solid line represents the fourth-order polynomial fitting. The black dashed line represents LISA's TTL requirement, and the gray dashed line represents the preliminary requirements for TianQin ground test. It can be observed that after passing through the imaging system, the TTL coupling in the system is significantly suppressed, meeting the required standards.}
    \label{fig11}
\end{figure}

\section{Conclusion}
TTL coupling is ubiquitous and any misalignment can lead to its occurrence. It is one of the main noise sources in space-based gravitational wave detection missions such as TianQin, LISA, and Taiji. Therefore, studying the formation mechanism of TTL coupling and the methods to suppress it are crucial for the development of ultra-high-precision space laser interferometers. One challenge in the experiments involving TTL coupling is that the longitudinal displacement error of the tilt actuator directly couples into the TTL calibration and interferes with the final results. To address this, we propose a novel tilt brake (APTA) to eliminate the longitudinal displacement error of the actuator, providing the interferometer system under test with a pure tilt drive. The APTA comprises two parts: a compact four-beam interferometer and a retroreflector tilt actuator. The APTA reflects the measurement beam in the interferometer under test and measures the three degrees of freedom information at the reflection point, thus enabling accurate TTL coupling calibration. The APTA features a compact and portable structure, making it applicable for TTL coupling experiments for future interferometers or any experiment requiring three degrees of freedom motion.

Additionally, we employ a classical confocal dual-lens imaging system to suppress TTL coupling. Experimental results show that after aligning the lenses and fine-tuning the detector position, the imaging system significantly suppresses TTL coupling. However, since off-the-shelf lenses were used and the actual layout of the optical platform in a space interferometer was not considered, future optimizations of the imaging system design based on the actual optical platform configuration are necessary.

During TTL coupling calibration, APTA still experiences high-order noise effects. We believe that misalignments during assembly and manufacturing errors in components such as the retroreflector are likely causes. Despite efforts to minimize errors during assembly, residual errors still interfere with the final calibration results. This study analyzes three error mechanisms within the APTA and employs data processing to minimize their impact on calibration. More research is required to investigate the error mechanisms within the APTA and optimize its structure and assembly process.

TTL coupling experiments show that without the imaging system and within a tilt range of $\pm 200$ $\mu$rad, the measured TTL coupling reaches up to 3 nm, with a coupling coefficient of $\pm 15$ $\mu$m/rad. Using the imaging system, the TTL coupling can be reduced to below 0.6 nm throughout the tilt range, with the coupling coefficient decreasing to within $\pm 5$ $\mu$m/rad.

Furthermore, within a small tilt range of $\pm 100$ $\mu$rad, the slope can be reduced to below 3 $\mu$m/rad, meeting the preliminary requirements of the TianQin mission.

\begin{backmatter}
\bmsection{Funding}
National Natural Science Foundation of China (12105375).National Key Research and Development Program of China (2022YFC2203901).

\bmsection{Acknowledgments}
We thank members of Prof. xxx for useful discussions. The authors acknowledge the experimental facility support from the PGMF (National Precision Gravity Measurement Facility).

\bmsection{Disclosures}
The authors declare no conflict of interest.

\bmsection{Data Availability Statement}
Data underlying the results presented in this paper are not publicly available at this time but may be obtained from the authors upon reasonable request. 

\bmsection{Supplemental document}
See Supplement 1 for supporting content. 

\end{backmatter}

\bibliography{references}

\providecommand{\noopsort}[1]{}\providecommand{\singleletter}[1]{#1}%
\begin{thebibliography}{10}
\newcommand{\enquote}[1]{``#1''}

\bibitem{gong2021concepts}
Y.~Gong, J.~Luo, and B.~Wang, \enquote{Concepts and status of chinese space gravitational wave detection projects,} {\protect\JournalTitle{Nature Astronomy}} \textbf{5}, 881--889 (2021).

\bibitem{luo2016tianqin}
J.~Luo, L.-S. Chen, H.-Z. Duan, \emph{et~al.}, \enquote{Tianqin: a space-borne gravitational wave detector,} {\protect\JournalTitle{Classical and Quantum Gravity}} \textbf{33}, 035010 (2016).

\bibitem{danzmann2003lisa}
K.~Danzmann and A.~R{\"u}diger, \enquote{Lisa technology—concept, status, prospects,} {\protect\JournalTitle{Classical and Quantum Gravity}} \textbf{20}, S1 (2003).

\bibitem{jennrich2009lisa}
O.~Jennrich, \enquote{Lisa technology and instrumentation,} {\protect\JournalTitle{Classical and Quantum Gravity}} \textbf{26}, 153001 (2009).

\bibitem{hu2017taiji}
W.-R. Hu and Y.-L. Wu, \enquote{The taiji program in space for gravitational wave physics and the nature of gravity,}  (2017).

\bibitem{luo2020brief}
Z.~Luo, Z.~Guo, G.~Jin, \emph{et~al.}, \enquote{A brief analysis to taiji: Science and technology,} {\protect\JournalTitle{Results in Physics}} \textbf{16}, 102918 (2020).

\bibitem{brzozowski2022lisa}
W.~Brzozowski, D.~Robertson, E.~Fitzsimons, \emph{et~al.}, \enquote{The lisa optical bench: an overview and engineering challenges,} {\protect\JournalTitle{Space Telescopes and Instrumentation 2022: Optical, Infrared, and Millimeter Wave}} \textbf{12180}, 237--253 (2022).

\bibitem{Paczkowski2022}
S.~Paczkowski, R.~Giusteri, M.~Hewitson, \emph{et~al.}, \enquote{Postprocessing subtraction of tilt-to-length noise in lisa,} {\protect\JournalTitle{Physical Review D}} \textbf{106}, 042005 (2022).

\bibitem{Armano2023}
J.~B. M.~Armano, H.~Audley and P.~Binetruy, \enquote{Tilt-to-length coupling in lisa pathfinder: A data analysis,} {\protect\JournalTitle{PHYSICAL REVIEW D}} \textbf{108}, 1--25 (2023).

\bibitem{Hartig2023}
G.~W. Marie-Sophie~Hartig, \enquote{Tilt-to-length coupling in lisa pathfinder: Analytical modeling,} {\protect\JournalTitle{PHYSICAL REVIEW D}} \textbf{108}, 1--13 (2023).

\bibitem{hartig2022geometric}
M.-S. Hartig, S.~Schuster, and G.~Wanner, \enquote{Geometric tilt-to-length coupling in precision interferometry: mechanisms and analytical descriptions,} {\protect\JournalTitle{Journal of Optics}} \textbf{24}, 065601 (2022).

\bibitem{hartig2023non}
M.-S. Hartig, S.~Schuster, G.~Heinzel, and G.~Wanner, \enquote{Non-geometric tilt-to-length coupling in precision interferometry: mechanisms and analytical descriptions,} {\protect\JournalTitle{Journal of Optics}} \textbf{25}, 055601 (2023).

\bibitem{Sonke2016}
G.~W. Sonke~Schuster, Michael~Trobs and G.~Heinzel, \enquote{Experimental demonstration of reduced tilt-to-length coupling by a two-lens imaging system,} {\protect\JournalTitle{Optics Express}} \textbf{24}, 10466--10475 (2016).

\bibitem{Tröbs2018}
M.~Tr{\"o}bs, S.~Schuster, M.~Lieser, \emph{et~al.}, \enquote{Reducing tilt-to-length coupling for the lisa test mass interferometer,} {\protect\JournalTitle{Classical and Quantum Gravity}} \textbf{35}, 105001 (2018).

\bibitem{Chwalla2016}
M.~Chwalla, K.~Danzmann, G.~F. Barranco, \emph{et~al.}, \enquote{Design and construction of an optical test bed for lisa imaging systems and tilt-to-length coupling,} {\protect\JournalTitle{Classical and Quantum Gravity}} \textbf{33}, 245015 (2016).

\bibitem{lee2021development}
Y.~Lee, \enquote{Development of an advanced tilt actuator for tilt-to-length coupling investigations,} Ph.D. thesis, Hannover: Institutionelles Repositorium der Leibniz Universit{\"a}t Hannover (2021).

\bibitem{lin2024compact}
X.~Lin, P.~Qiu, Y.~Liang, and H.~Yan, \enquote{Compact auto-aligning interferometers with picometer precision,} {\protect\JournalTitle{Applied Optics}} \textbf{63}, 3910--3915 (2024).

\bibitem{lin2023construction}
X.~Lin, H.~Yan, Y.~Ma, and Z.~Zhou, \enquote{A construction method of the quasi-monolithic compact interferometer based on uv-adhesive bonding,} {\protect\JournalTitle{Review of Scientific Instruments}} \textbf{94} (2023).

\bibitem{schuster2017tilt}
S.~Schuster, \enquote{Tilt-to-length coupling and diffraction aspects in satellite interferometry,} Ph.D. thesis, Hannover: Institutionelles Repositorium der Leibniz Universit{\"a}t Hannover (2021).

\bibitem{wanner2015brief}
G.~Wanner, S.~Schuster, M.~Tr{\"o}bs, and G.~Heinzel, \enquote{A brief comparison of optical pathlength difference and various definitions for the interferometric phase,} in \emph{Journal of Physics: Conference Series,}  vol. 610 (IOP Publishing, 2015), p. 012043.

\bibitem{Chwalla2020}
M.~Chwalla, K.~Danzmann, M.~D. {\'A}lvarez, \emph{et~al.}, \enquote{Optical suppression of tilt-to-length coupling in the lisa long-arm interferometer,} {\protect\JournalTitle{Physical Review Applied}} \textbf{14}, 014030 (2020).

\bibitem{wei2010}
W.~Ruofei, \enquote{Study on alignment method of a laser synthetic wavelength nanomeasurement interferometer,} Master's thesis, Zhejiang Sci-Tech University (2010).

\end{thebibliography}

\end{document}